# Temperature dependence of spin polarization in ferromagnetic metals using lateral spin valves

Estitxu Villamor,[1] Miren Isasa,[1] Luis E. Hueso,[1,2] and Fèlix Casanova[1,2] [*]

[1]CIC nanoGUNE, 20018 Donostia-San Sebastian, Basque Country (Spain)

[2]IKERBASQUE, Basque Foundation for Science, 48011 Bilbao, Basque Country (Spain)

[*] E-mail: f.casanova@nanogune.eu

Spin injection properties of ferromagnetic metals are studied and are compared by using highly reproducible cobalt/copper and permalloy/copper lateral spin valves (LSVs) with transparent contacts, fabricated with a careful control of the interface and the purity of copper. Spin polarization of permalloy and cobalt are obtained as a function of temperature. Analysis of the temperature dependence of both the spin polarization and the conductivity of permalloy confirms that the two-channel model for ferromagnetic metals is valid to define the current spin polarization and shows that a correction factor of ~2 is needed for the values obtained by LSV experiments. The spin transport properties of copper, which also are studied as a function of temperature, are not affected by the used ferromagnetic material. The low-temperature maximum in the spin-diffusion length of copper is attributed to the presence of diluted magnetic impurities intrinsic from the copper.

## I. INTRODUCTION

During the past years, lateral spin valves (LSVs) have gained increasing attention in the field of *spintronics* [1], which aims at taking advantage of the spin degree of freedom for electronics performance. Being able to decouple the spin current from the charge current, these devices are interesting due to their potential application to information technology as well as from a fundamental point of view. LSVs consist of two ferromagnetic (FM) electrodes bridged by a nonmagnetic (NM) channel (see Fig. 1(a)), which allow the electrical injection of a pure spin current from one of the FMs into the NM (and its detection with the other FM) due to their nonlocal geometry [2-21]. Eliminating any spurious effects coming from the charge, they provide an effective way for studying the spin transport mechanisms inside a NM material (including metals [2-9], semiconductors [10], superconductors [11] and graphene [12]), as well as the spin injection in the FM/NM system [13-21].

Despite the large number of reports employing LSVs, the dispersion in the obtained data is fairly high in literature. As an example, Table I shows values of the spin diffusion length of copper (Cu), $\lambda_{Cu}$, and the spin polarizations of permalloy (Py) and cobalt (Co), $\alpha_{Py}$ and $\alpha_{Co}$, obtained from references using Py/Cu and Co/Cu LSVs with transparent interfaces. The main reason for such dispersion lies in the irreproducibility of the fabrication of the devices [4,5,21,22], due to uncontrollable factors relevant at the nanoscale, which can lead to different results. For instance, a small variation in the interface quality can induce a large change in the effective values of $\alpha_{Py}$ and $\alpha_{Co}$, deduced from the one-dimensional (1D) spin-diffusion model [23,24]. Since the spin-flip scattering in metals is governed by the Elliott-Yafet mechanism [25,26], $\lambda_{Cu}$ should change linearly with the inverse of the Cu channel resistivity, $1/\rho_{Cu}$ [9,27]. However, the dispersion in $\lambda_{Cu}$ is too large to solely be explained by the difference in $\rho_{Cu}$ (see Table I). Magnetic impurities at the NM channel, which strongly decrease the spin diffusion length of a NM material, are the most likely reason for such dispersion [4,8].

In this paper we show that, as a result of a careful optimization, we are able to fabricate reproducible LSVs with transparent interfaces, i.e., the obtained spin signal for a given material, dimensions and interface treatment are always the same, yielding consistent values of $\alpha_{FM}$ and $\lambda_{Cu}$. These values can thus be considered as a reference in LSV experiments. This allows a systematic analysis of Py/Cu and Co/Cu LSV systems, by studying the role both FM materials play on the spin transport in Cu and comparing their spin injection properties. Moreover, the current spin polarization of Py and Co as a function of temperature is reported for LSVs with transparent interfaces. The simultaneous analysis of the current spin polarization and the conductivity of Py based on the standard two-channel model [28,29], allows us to correct a systematic underestimation of $\alpha_{Py}$ derived from the LSV experiments.

## II. EXPERIMENTAL DETAILS

Samples were fabricated by two consecutive electron-beam lithography and lift-off steps. In the first step, FM electrodes were patterned in polymethylmethacrylate resist on top of SiO$_2$(150 nm)/Si substrates and 35 nm of Py or Co were deposited in an UHV electron-beam evaporator (base pressure ≤1·10$^{-8}$ mbar). Different widths of FM electrodes were chosen, ~110 and ~150 nm, in order to obtain different switching magnetic fields. In the second step, the NM channel with a width of ~170 nm and contact pads were patterned and Cu was thermally evaporated with the same base pressure $\leq 1\cdot 10^{-8}$ mbar. Prior to the Cu deposition, in order to ensure a transparent interface, the surface of the FM electrodes was cleaned from oxidation and resist left-overs by Ar-ion milling. This process was optimized as described below. Each sample contains up to ten different LSVs with an edge-to-edge separation distance ($L$) between FM electrodes from 200 to 3500 nm. Figure 1(a) is a scanning electron microscopy (SEM) image of a device.

Non-local measurements were performed in a liquid He cryostat (with applied magnetic field $H$ and temperatures ranging from 10 to 300 K) using a "dc reversal" technique [13]. The voltage $V$, normalized to the absolute value of the applied current $I$, is defined as the non-local resistance $R_{NL} = V/|I|$ (see Fig. 1 (a) for a scheme of the measurement). This magnitude is positive (negative) when the magnetization of the electrodes is parallel (antiparallel), depending on the value of $H$. The difference $\Delta R_{NL}$ between the positive and the negative values of $R_{NL}$ is called spin signal (see Fig. 1 (b)), which is proportional to the spin accumulation at the detector. Applying the 1D spin-diffusion model to our geometry, the following expression is obtained for the spin signal [7,14,23]:

$$\Delta R_{NL} = \frac{2\alpha_{FM}^2 R_{NM}}{\left(2+\frac{R_{NM}}{R_{FM}}\right)^2 e^{L/\lambda_{NM}} - \left(\frac{R_{NM}}{R_{FM}}\right)^2 e^{-L/\lambda_{NM}}}, \qquad (1)$$

where $\alpha_{FM}$ is the spin polarization of the FM, $R_{NM} = 2\lambda_{NM}\rho_{NM}/t_{NM}w_{NM}$ and $R_{FM} = 2\lambda_{FM}\rho_{FM}/(1-\alpha_{FM}^2)w_{FM}w_{NM}$ are the spin resistances, $\lambda_{NM,FM}$ are the spin diffusion lengths, $\rho_{NM,FM}$ are the resistivities and $w_{NM,FM}$ are the widths of the NM and the FM, respectively, and $t_{NM}$ is the thickness of the NM. $\rho_{Cu}$ as a function of temperature is obtained by measuring the resistance of Cu for every $L$ and performing a linear regression for each sample, whereas $\rho_{Co}$ (= 11.5 μΩ cm at 10 K) and $\rho_{Py}$ (= 22.4 μΩ cm at 10 K) are obtained separately as a function of temperature in two different devices, where the FM materials were deposited under the same nominal conditions as the FM electrodes of the LSVs. We use $\lambda_{Py} = 5$ nm [30] and $\lambda_{Co} = 36$ nm [30] at 10 K and consider a temperature dependence of the form $\lambda_{FM} = const./\rho_{FM}$ as deduced from the Valet-Fert theory and the free electron model [30,31]. The previous scaling relation is valid for all temperature ranges because, even if electron-magnon scattering is present above 100 K, it is not an efficient spin-lattice relaxation mechanism [31,32]. Geometrical parameters are measured by SEM for each device. The values of $\alpha_{FM}$ and $\lambda_{NM}$ are obtained by fitting $\Delta R_{NL}$ as a function of $L$ to Eq. (1).

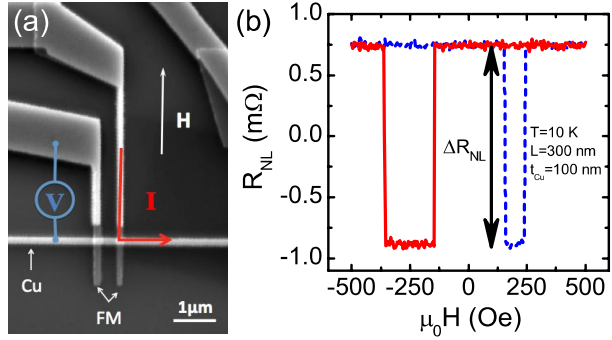

Figure 1: (a) SEM image of a typical lateral spin valve. The non-local measurement configuration, materials and the direction of the applied magnetic field $H$ are shown. (b) Non-local resistance, measured at 10 K, for a Py/Cu lateral spin valve with $t_{Cu}=$ 100 nm and $L$=300 nm. Solid red (dashed blue) line indicates the decreasing (increasing) direction of the magnetic field. Spin signal is tagged as $\Delta R_{NL}$. No background is subtracted from the measured data.

The spin injection efficiency at a transparent junction is very sensitive to the interface quality [13,17,21]. In order to optimize the interface cleaning process, $\alpha_{FM}$ was obtained for different samples in which the Ar-ion milling time was systematically changed. The other milling parameters were kept constant, with an Ar flow of 15 standard cubic centimeters per minute, an acceleration voltage of 50 V, a beam current of 50 mA and a beam voltage of 300 V. The inset in Figure 2 shows the spin polarization of Py as a function of the Ar-ion milling time, where $\alpha_{Py}$ increases with time, as expected from the interface cleaning process in which the resist left-overs and the oxide are being removed. After a maximum ($\alpha_{Py} = 0.39 \pm 0.01$) is obtained for a milling time of 30 s, the value of $\alpha_{Py}$ starts to decrease for longer times. This can be understood by the fact that the milling process increases the roughness of the FM surface (as checked by atomic force microscope) once it is completely cleaned, leading to a rougher interface with reduced spin injection efficiency. It is worth noting that the Ar-ion milling was not performed *in-situ*, i.e. the vacuum was broken for entering the sample into the Cu evaporation chamber. However, this is shown not to be crucial for obtaining a transparent interface with the highest spin polarization.

The interface resistance was measured in some of the optimized devices, where a cross-shaped junction was added to the original design. The value of the measured resistance was negative in all junctions. This is an artifact which occurs when the resistance of the electrodes is on the order or higher than the interface resistance [33,34], due to an inhomogeneous current distribution in this geometry. We can thus estimate the value of the interface resistance multiplied by the area to be ≤ 1 × 10$^{-3}$ Ωμm$^2$, confirming that we are indeed in the transparent regime [24].

## III. RESULTS AND DISCUSSION
### A. Reproducibility of the lateral spin valves

As discussed in the Introduction, the main sources of irreproducibility are both the presence of magnetic impurities and the quality of the interface. In the present paper, the former is avoided by using the two-step lithography technique as opposed to the two-angle shadow evaporation technique, thus, avoiding cross-contamination between FM and NM [8]. The latter is solved by optimizing a protocol to obtain the interface with the same good quality. As an example, Fig. 2 displays the measured $\Delta R_{NL}$ as a function of $L$ at 10 K for four different Cu-based samples, two with Py electrodes and two with Co electrodes. Since the value of $\lambda_{Cu}$ is influenced by the grain boundary scattering [9], the Cu channel dimensions are kept constant (with a thickness of ~100 nm). The results match perfectly for the two pairs of samples with the same FM/NM combination. In addition, since $\Delta R_{NL}$ decays nearly exponentially with $L$ (see Eq. (1)), the slope in the semilogarithmic plot is essentially $\lambda_{Cu}$, remaining the same for all four samples. Furthermore, the clear shift in $\Delta R_{NL}$ for samples with different FM materials is caused by their different spin injection efficiency $\gamma = \left(\frac{2\alpha_{FM}\lambda_{FM}\rho_{FM}}{1-\alpha_{FM}^2}\right)^2$ [15], which is directly related to the intrinsic properties of the FM metal and is an important contribution to Eq. (1). The consistent results shown in Fig. 2, which have been reproduced for virtually all samples we have fabricated (more than 20), allow us to compare properties between different samples, a long-standing problem in this type of device [4,5,21,22].

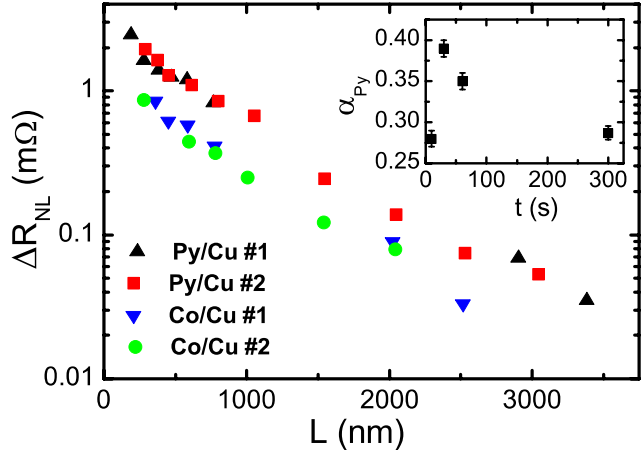

Figure 2: Spin signal as a function of the distance $L$ between FM electrodes measured at 10 K for 4 different samples with a Cu channel and Py or Co as a FM. Inset: Spin polarization of Py as a function of the Ar-ion milling time. The error bars correspond to the error from the fitting of Eq. (1).

## B. Spin transport in Cu using different FM metals

Next, we study in more detail the role that Py and Co play in the spin transport of Cu and the spin injection properties of both FM materials. With this purpose, we obtain $\lambda_{Cu}$ and $\alpha_{FM}$ as a function of temperature in two new samples with identical characteristics, where ~70 nm of Cu were deposited as the NM channel, and the FM electrodes were made of Py in one case and of Co in the other.

The values and the temperature dependence of $\lambda_{Cu}$, shown in Fig. 3(a), are the same for both samples containing Py and Co electrodes ($\lambda_{Cu}$ = 860 ± 20 nm and 820 ± 90 nm at 10 K for Co and Py, respectively). This is consistent with the fact that they show the same resistivity $\rho_{Cu}$ at all temperatures ($\rho_{Cu}$ ~ 1.6 μΩ cm at 10 K, see the inset of Fig. 3(a)), since $\lambda_{Cu}$ is basically proportional to $1/\rho_{Cu}$ [9]. This good agreement evidences that the use of different FM electrodes does not influence the spin transport properties of the NM channel. The obtained values of $\lambda_{Cu}$ are among the highest reported in LSV experiments, given the dimensions and the $\rho_{Cu}$ of the channel (see Table 1), further suggesting that the purity of the Cu channel is not affected by the fabrication process.

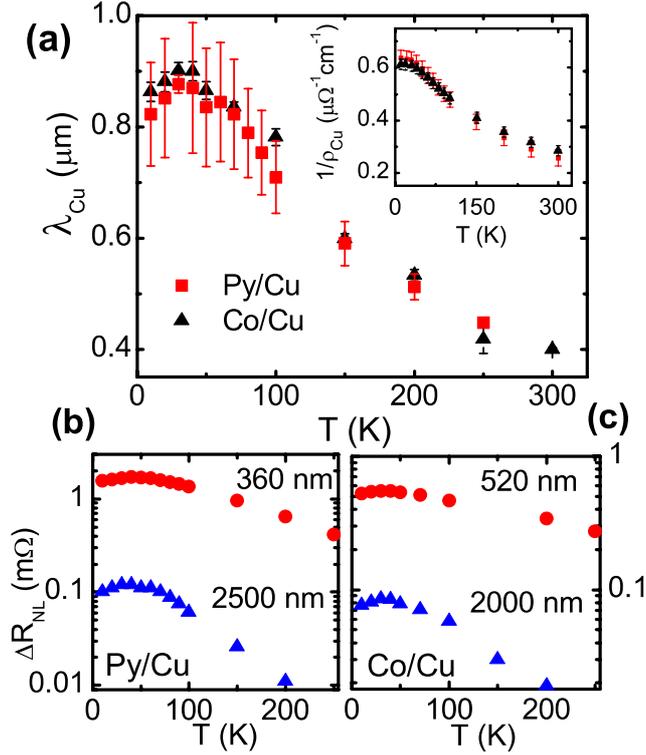

Figure 3: (a) Spin diffusion length of Cu as a function of temperature using Py (red squares) and Co (black triangles) electrodes. Error bars correspond to the error from the fitting of Eq. (1). A maximum is found at 30 K. Inset: Inverse of the resistivity of Cu as a function of temperature using Py (red squares) and Co (black triangles) electrodes. Error bars correspond to the error from the linear regression used to obtain the resistivity. (b) Spin signal as a function of temperature in two Py/Cu LSVs with L = 360 nm (red dots) and L = 2500 nm (blue triangles). A maximum is observed at 30 K. (c) Spin signal as a function of temperature in two Co/Cu LSVs with L = 520 nm (red dots) and L = 2000 nm (blue triangles). A maximum is observed at 30 K.

Despite the saturation of $1/\rho_{Cu}$ at low temperatures, a maximum is found in $\lambda_{Cu}$ for both samples, which previously has been reported for Cu- and Ag-based LSVs [5-9,14]. The maximum in $\lambda_{Cu}$ arises from a maximum in $\Delta R_{NL}$ as a function of temperature, which also occurs at 30 K in both Py/Cu and Co/Cu LSVs for any *L*, as shown in Fig. 3(b) and 3(c). The origin of such a maximum has been suggested to arise from an additional spin-flip scattering at the surface of the NM channel [5-7]. This extra spin-flip scattering at the surface has been attributed to the presence of magnetic impurities [8,9], which can appear as a consequence of using a two-angle shadow evaporation process [8]. In our case, the presence of magnetic impurities at the surface of the NM channel is unlikely due to the two-step fabrication process employed, and thus they must be located at the bulk. Interdiffusion of the FM material near the interface could be a possible source of magnetic impurities at the NM channel. In this case, the effective spin injection should also be affected, leading to the observation of a clear maximum in the temperature dependence of the spin polarization [35]. However, this is not observed in our samples (see Fig. 4). In addition, changing the FM material should also change the position of the maximum in $\lambda_{Cu}$ [35], but, in the present study, the shape and position (30 K) of the maximum in $\lambda_{Cu}$ is independent of the used FM material, as shown in Fig. 3(a). These two observations rule out the presence of interdiffusion and confirm the previous evidence that, with our fabrication process, the NM channel is not contaminated in any way by the FM used in the electrodes. Therefore, the magnetic impurities at the bulk of the Cu channel, responsible for the maximum in $\lambda_{Cu}$, must be introduced during the evaporation process, probably from the original Cu source. This result is in agreement with recent observations in which the origin of the maximum in $\lambda_{Cu}$ is attributed to magnetic impurities at the bulk of the NM channel coming from different sources [35].

## C. Spin injection properties of Py and Co

The values and temperature dependence of $\alpha_{Py}$ and $\alpha_{Co}$ are shown in Fig. 4. The values at 10 K are $\alpha_{Py} = 0.38 \pm 0.01$ and $\alpha_{Co} = 0.118 \pm 0.001$. The obtained $\alpha_{Py}$ is among the highest values reported in LSV experiments (see Table 1), but is usually lower (down to half) than the values obtained by other methods (0.47-0.75) [30,36-38]. $\alpha_{Co}$ is also on the same order as the highest reported values in LSV experiments with transparent interfaces (see Table 1) and, in this case, much lower (three to five times) than the values obtained by other methods (0.36-0.52) [31,37,39]. A possible explanation for the dramatic difference in Co is the uncertainty on the value of $\lambda_{Co}$, a parameter used in the fitting to Eq. (1). Co has been reported to have $\lambda_{Co} \sim$ 40-60 nm [31,39,40], a value questioned for being much longer than those of the rest of the FM materials [27]. Since the values of $\alpha_{FM}$ and $\lambda_{FM}$ are coupled in Eq. (1) [3] and it is not possible to obtain them independently, an overestimation of $\lambda_{Co}$ would lower the fitted value of $\alpha_{Co}$. The quality of the Co/Cu interface could be another reason for the low $\alpha_{Co}$. Due to the natural oxidation of Co, a spin-independent interface resistance might be created between the two metals, which would act as an additional spin-flip scatterer [3]. Since this additional contribution is not taken into account in Eq. (1), it would reduce the fitted value of $\alpha_{Co}$.

At this point, it should be mentioned that, even if the obtained $\alpha_{Co}$ is three times smaller than $\alpha_{Py}$, the product $\alpha_{FM}\lambda_{FM}$ is twice as large for Co than for Py. The only other contribution to the spin injection efficiency $\gamma$ is the electrical resistivity of the FM metal. Since $\rho_{FM}$ is lower for Co, the backflow of the spin current is higher [15,16], making the spin injection less effective, as observed in the shift in $\Delta R_{NL}$ (Fig. 2).

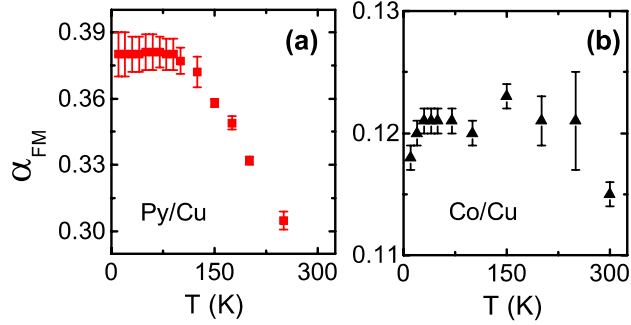

Figure 4: Spin polarization as a function of temperature for (a) Py and (b) Co. The error bars correspond to the error from the fitting of Eq. (1).

### D. Temperature dependence of the spin polarization of Co and Py

In order to analyze the temperature dependence of $\alpha_{FM}$ extracted from the fitting of our measurements (Fig. 4), we will first note that it is an independent measurement fully decoupled from the temperature dependence of $\lambda_{FM}$, which is directly obtained from the FM resistivity [9]. It is also worth noting that the magnitude obtained from LSV experiments with transparent contacts corresponds to the current spin polarization [3]:

$$\alpha_{FM} = \frac{\sigma_\uparrow - \sigma_\downarrow}{\sigma_\uparrow + \sigma_\downarrow}, \quad (2)$$

where $\sigma_\uparrow$ and $\sigma_\downarrow$ are the spin-dependent conductivities introduced in Mott's two-channel model [29] and further developed by Fert and Campbell [28] in the study of electronic transport in pure ferromagnetic materials as well as in ferromagnetic alloys. However, the temperature dependence of the current spin polarization of FM metals has only recently been reported and analyzed [38], and to our knowledge, values obtained using LSV experiments have not been studied as a function of temperature before. In the case of Co, the decay in $\alpha_{Co}$ is negligible up to 300 K (Fig. 4 (b)), which is expected from previous experiments [31]. In the case of Py, a clear decay in $\alpha_{Py}$ is observed with temperature (Fig. 4 (a)), which will allow us to analyze this dependence.

The spin-dependent conductivities can be written as:

$$\sigma_\uparrow = \frac{\rho_\downarrow + 2\rho_{\uparrow\downarrow}}{\rho_\uparrow \rho_\downarrow + \rho_{\uparrow\downarrow}(\rho_\uparrow + \rho_\downarrow)} \;\; ; \;\; \sigma_\downarrow = \frac{\rho_\uparrow + 2\rho_{\uparrow\downarrow}}{\rho_\uparrow \rho_\downarrow + \rho_{\uparrow\downarrow}(\rho_\uparrow + \rho_\downarrow)} \qquad (3)$$

where $\rho_\uparrow$ and $\rho_\downarrow$ are the resistivities for spin up and down channels and $\rho_{\uparrow\downarrow}$ is the spin mixing resistivity, which is a measure of the momentum transfer between the two channels by spin mixing scatterings (basically caused by electron-magnon scattering) [28,32]. The total conductivity through a FM material, thus, has the form:

$$\sigma_{FM} = \sigma_\uparrow + \sigma_\downarrow = \frac{\rho_\uparrow + \rho_\downarrow + 4\rho_{\uparrow\downarrow}}{\rho_\uparrow \rho_\downarrow + \rho_{\uparrow\downarrow}(\rho_\uparrow + \rho_\downarrow)} \qquad (4)$$

and $\alpha_{FM}$ can be represented as a function of each spin-dependent resistivity:

$$\alpha_{FM} = \frac{\rho_\downarrow - \rho_\uparrow}{\rho_\uparrow + \rho_\downarrow + 4\rho_{\uparrow\downarrow}} . \qquad (5)$$

The temperature dependence of the three spin-dependent resistivities is modeled by considering that $\rho_i = \rho_{0i} + A_i T^2$ ($i = \uparrow, \downarrow$ and $\uparrow\downarrow$), where the term $\rho_{0i}$ accounts for spin flip-scattering due to impurities [28] and the temperature dependence comes from phonon and magnon scattering [38]. Hence, we can see from Eq. (5) that the increase of $\rho_i$ with temperature will lower $\alpha_{FM}$.

In order to explain the experimental temperature decay of $\alpha_{Py}$ with Eq. (5), coefficients $\rho_{0i}$ and $A_i$ are calculated. Assuming that the spin mixing resistivity, and thus $\rho_{0\uparrow\downarrow}$, is zero at low temperatures [32,38], a ratio between $\rho_{0\uparrow}$ and $\rho_{0\downarrow}$ can be obtained from the low temperature values of $\sigma_\uparrow$ and $\sigma_\downarrow$. The values of $\sigma_\uparrow$ and $\sigma_\downarrow$ of Py are calculated from the experimental values of $\alpha_{Py}$ and $\sigma_{Py}$ using Eq. (2) and (4). The ratio $\rho_{0\uparrow}/\rho_{0\downarrow} = 2.2$ that we obtain is lower than the values between 6 and 20 previously reported [28,38]. Next, $A_i$ coefficients have been fixed as $A_\uparrow = A_\downarrow = A_{\uparrow\downarrow}$, following Ref. 38. The conductivity of Py as a function of temperature has been plotted in Fig. 5(a) and fitted to Eq. (4) (red solid line), obtaining $\rho_{0\uparrow} = (3.230 \pm 0.001) \cdot 10^{-7}$ $\Omega \cdot m$ and $A_i = (1.96 \pm 0.01) \cdot 10^{-12}$ $\Omega \cdot m/K^2$. According to the model above, we should be able to reproduce the temperature dependence of $\alpha_{Py}$ by introducing these parameters into Eq. (5). However, the obtained curve (red solid line in Fig. 5(b)) does not reproduce well the experimental temperature dependence. Alternatively, we have fitted the experimental values of $\alpha_{Py}$ directly to Eq. (5) (blue dashed line in Fig 5(b)) by fixing $\rho_{0\uparrow} = 3.230 \cdot 10^{-7}$ $\Omega \cdot m$, obtaining in this case $A_i = (0.60 \pm 0.02) \cdot 10^{-12}$

Ω·m/K$^2$. In turn, this $A_i$ value cannot reproduce the experimental values of $\sigma_{Py}$ (blue dashed line in Fig 5(a)).

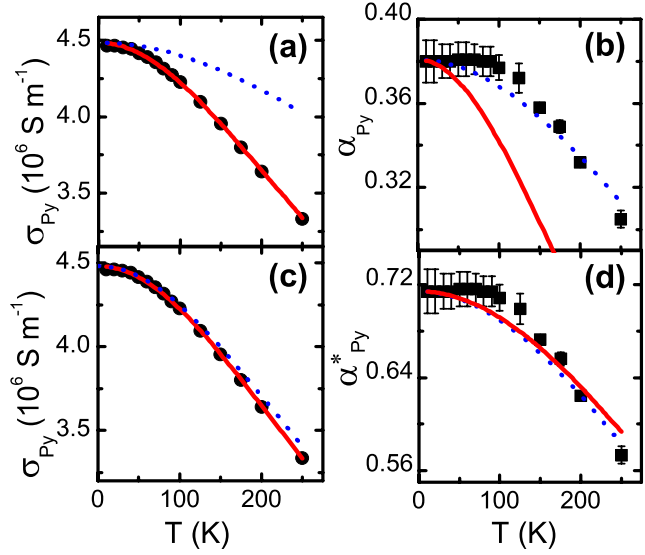

Figure 5: (a) Conductivity of Py as a function of temperature (black dots). Red solid line is a direct fit of the data to Eq. (4) and blue dashed line is the representation of Eq. (4) with the $\rho_{0\uparrow}$ and $A_i$ parameters obtained from the fitting of the data in (b) to Eq. (5). (b) Spin polarization of Py as a function of temperature (black squares). Blue dashed line is a direct fit of the data to Eq. (5) and red solid line is the representation of Eq. (5) with the $\rho_{0\uparrow}$ and $A_i$ parameters obtained from the fitting of the data in (a) to Eq. (4). (c) Conductivity of Py as a function of temperature (black dots). Red solid line is a direct fit of the data to Eq. (4) and blue dashed line is the representation of Eq. (4) with the $\rho_{0\uparrow}$ and $A_i$ parameters obtained from the fitting of the data in (d) to Eq. (5). (d) Corrected spin polarization of Py as a function of temperature (black squares). Blue dashed line is a direct fit of the data to Eq. (5) and red solid line is the representation of Eq. (5) with the $\rho_{0\uparrow}$ and $A_i$ parameters obtained from the fitting of the data in (c) to Eq. (4).

The mismatch between the red solid and the blue dashed lines in both Figs. 5(a) and 5(b) evidences that the model assumed cannot simultaneously describe our two sets of independent data ($\alpha_{Py}$ and $\sigma_{Py}$). Considering the validity of the model to explain the temperature dependence of the current spin polarization in a previous study [38], it is more plausible to suppose that the origin of the disagreement comes from the data sets. In particular, the obtained low values of $\alpha_{Py}$ in LSV experiments compared to other experiments (*i.e.* giant magnetoresistance [27,28] or spin-wave Doppler [38]), suggest an underestimation of our $\alpha_{Py}$ data, which can be corrected by introducing a multiplying factor to the experimental $\alpha_{Py}$ values. A factor of 1.88 is found to give the best agreement between our two data sets ($\alpha_{Py}$ and $\sigma_{Py}$) and the model above, as well as a much closer match between our $\alpha_{Py}$ value and the ones obtained by other methods [30,36,38]. With such a correction, the ratio $\rho_{0\uparrow}/\rho_{0\downarrow}$ has a value of 6, now within the range reported in Refs. 28 and 38. The parameters obtained from the fitting of $\sigma_{Py}$ to Eq. (4) (red solid curve in Fig. 5(c)) are $\rho_{0\uparrow}= (2.603 \pm 0.001) \cdot 10^{-7}$ $\Omega \cdot$m and $A_i = (1.09 \pm 0.01) \cdot 10^{-12}$ $\Omega \cdot$m/K$^2$, which being introduced into Eq. (5), perfectly reproduce the experimental curve of $\alpha^*_{Py}$ (red solid curve in Fig. 5(d)). Similar to what we have performed for $\alpha_{Py}$, we have fitted the values of $\alpha^*_{Py}$ directly to Eq. (5) (blue dashed line in Fig 5(d)) by fixing $\rho_{0\uparrow} = 2.603 \cdot 10^{-7}$ $\Omega \cdot$m, and obtaining $A_i = (0.99 \pm 0.02) \cdot 10^{-12}$ $\Omega \cdot$m/K$^2$. This value of $A_i$ is now in excellent agreement with the previous one, and reproduces with high accuracy the experimental values of $\sigma_{Py}$ (blue dashed line in Fig 5(c)).

There are several possible reasons which are not mutually exclusive, for the underestimation of the obtained $\alpha_{Py}$. The first one could be the choice of a wrong injection area in the expression of $R_{FM}$ present in Eq. (1). If we consider that the side surfaces of the FM electrodes are also in contact with the Cu channel, the correct expression would be $R_{FM} = 2\lambda_{FM}\rho_{FM}/(1-\alpha^2_{FM})(w_{FM}w_{NM} + 2t_{FM}w_{FM})$. By introducing such a correction into Eq. (1) $\alpha_{Py}$ increases from 0.38 to 0.49 at 10 K. Another possible reason for such underestimation could be the 1D approximation of the spin-diffusion model [23] used in LSV [7,14] to derive Eq. (1). Indeed, it has already been reported [4] that such an approximation does not consider the "intermediate" region of the NM metal above the FM/NM interface, which causes spins to flip before they diffuse through the NM channel, and even to flow back to the FM electrodes, underestimating the fitted $\alpha_{FM}$ value. Similarly, Niimi *et al.* [41] have recently analyzed LSV data using a 3D finite element model based on an extension of the Valet-Fert formalism, where they observe an increase in the fitted $\alpha_{Py}$ from 1D to 3D.

## IV. CONCLUSIONS

To summarize, we succeeded in obtaining reproducible LSV devices with transparent contacts due to an optimized nanofabrication method based on a two-step lithography. This allows us to compare properties between different samples, a long-standing problem in this type of device. Regarding the spin transport properties in Cu, the values and temperature dependence of $\lambda_{Cu}$ are the same regardless of the FM material used, including a maximum at 30 K. This result shows that no contamination from the FM material into the NM channel is induced, enhancing the spin transport, and evidences that the spin-flip scattering sources responsible for the observed maximum are intrinsic magnetic impurities present in the Cu. The electrical spin injection from both Py and Co is also compared, clearly observing a decreased spin injection with the latter one, caused by its lower electrical resistivity. The experimental spin polarizations of both FM materials are among the highest reported in LSV experiments, even though they are systematically lower than those obtained by other methods, and their temperature dependences are reported. For the case of Py, the comparison of the temperature dependence of the spin polarization with the conductivity agrees well with the prediction given by the standard two-channel model, but a correction factor of ~2 to the spin polarization is detected. Our analysis thus confirms the substantial underestimation of the spin polarization in LSV experiments and identifies several contributions to this mismatch.

## ACKNOWLEDGMENTS

This work was supported by the European Union 7th Framework Programme under the Marie Curie Actions (PIRG06-GA-2009-256470) and the European Research Council (Grant 257654-SPINTROS), by the Spanish Ministry of Science and Education under Project No. MAT2012-37638 and by the Basque Government under Project No. PI2011-1. E. V. and M. I. thank the Basque Government for a PhD fellowship (Grants No. BFI-2010-163 and No. BFI-2011-106).

Table I: Resistivity and spin diffusion length of Cu and spin polarizations of Py and Co extracted from several references using Py/Cu and Co/Cu LSVs with transparent interfaces.

| T (K) | $\rho_{Cu}$ (μΩ cm) | $\lambda_{Cu}$ (nm) | $\alpha_{Py}$ | $\alpha_{Co}$ | Reference |
|---|---|---|---|---|---|
| 4.2 | 2.8 | 1000 | 0.2 | | 3 |
| 10 | 1.36 | 200 | | 0.074 | 13 |
| 77 | 1.14 | 1500 | 0.25 | | 17 |
| 10 | 0.69 | 1000 | 0.58 | | 7 |
| 4.2 | 1.67 | 395 | 0.29 | | 14 |
| 4.2 | -- | 190-260 | | 0.1 | 18 |
| 4.2 | -- | 100-400 | 0.15-0.4 | | 4 |
| 4.2 | 1.5 | 460 | 0.21 | | 8 |
| 80 | 1.2 | 1300 | 0.35 | | 19 |
| 4.2 | 4 | 400 | 0.33 | | 20 |
| 10 | 1.18 | 1020 | 0.4 | | 9 |
| 10 | 1.6 | 820,860 | 0.38 | 0.12 | This work |